\documentclass[singlespacing]{elsart}
\usepackage{amssymb,amsmath}
\usepackage{graphicx}
\usepackage{epsfig}
\usepackage{color}
\usepackage{subfigure}

\newtheorem{theorem}{Theorem}[section]
\newtheorem{lemma}[theorem]{Lemma}

\newtheorem{proposition}[theorem]{Proposition}
\newtheorem{corollary}[theorem]{Corollary}

\newtheorem{remark}[theorem]{Remark}
\newtheorem{definition}[theorem]{Definition}
\DeclareMathOperator{\trace}{trace}

\def\0{{\bf 0}}

\def\p{{\mathbf{p}}}
\def\q{{\mathbf{q}}}

\def\s{{\mathbf{s}}}

\def\u{{\bf u}}

\def \beq{\begin{equation}}
\def \eeq{\end{equation}}

\begin{document}

\begin{frontmatter}

\title{Escape dynamics in collinear atomic-like three mass point  systems}

\author{Daniel Pa\c sca}
\ead{dpasca@uoradea.ro}
\address{Department of Mathematics
and Informatics, University of Oradea, University Street, No. 1, RO-410087,
Oradea, Romania}

\author{Manuele Santoprete}
\ead{msantoprete@wlu.ca}
\address{Department of Mathematics,
Wilfrid Laurier University,
Waterloo, N2L 3C5, Ontario, Canada}

\author{Cristina Stoica \corauthref{cor1}}
\ead{cstoica@wlu.ca}
\corauth[cor1]{corresponding author}

\address{Department of Mathematics,
Wilfrid Laurier University,
Waterloo, N2L 3C5, Ontario, Canada}

\begin{abstract}

The present paper studies the escape mechanism in  collinear  three point mass systems with  small-range-repulsive/large-range-attractive pairwise-interaction. 
Specifically, we focus on systems with non-negative total energy.

We show that  on the zero energy level set, most of the orbits lead to binary escape configurations and the set of initial conditions leading to  escape configurations where all three separations infinitely increase as $t \to \infty$ has zero Lebesque measure. We also give numerical evidence of the existence of a periodic orbit for the case when the two outer masses are equal. 




For positive energies, we prove that the set of initial conditions leading to  escape configurations where all three separations infinitely increase as $t \to \infty$ has positive Lebesque measure.

{\textit {Keywords:} linear three point mass systems,  Lennard-Jones potential, positive total energy, near escape dynamics, infinity manifold, dissociation configurations}

 \end{abstract}

\end{frontmatter}

\section{Introduction}

In computational chemistry  and molecular dynamics it is common to use 
mathematically  simple models  to   describe the interaction between a pair of  atoms or molecules (see, for instance, \cite{Frenkel} and   \cite{Lewars}). Set within the framework of the classical  $n$-body problem, these models ignore  quantum effects for the gain of  significantly less computation time in numerical simulations. 
%

The most commonly encountered molecular  potential is the Lennard-Jones potential (or the 6-12 potential), which in given by the sum of a repulsive term at present at short ranges due to overlapping electron orbitals, and a long range  attractive   term (the van der Waals force, or dispersion force). The history of the Lennard-Jones potential and the motivation of its use for modeling molecules can be found in \cite{Brush}. The Lennard--Jones potential is often used to describe the properties of gases,  it is particularly accurate for noble gas atoms and it is   a good approximation  for neutral atoms and molecules.



In this paper we study a system formed by three mass points (atoms) confined to move on a line, with pairwise interaction given by an attractive-repusive  potential of the form:
\begin{equation}\label{LJ}
W(r):=- 4 \epsilon \left( \left(\frac{r_0}{r}\right)^{a}- \left(\frac{r_0}{r}\right)^{b},  \right)
\end{equation}
where $\epsilon$ and  $r_0$ are given real positive parameters, and  $a, b \in \mathbb{R}$    such that $4<a< b-1.$    
  Note that for $a=6$ and $b=12$ we retrieve the standard molecular  Lennard-Jones potential. 
 Thus one could think of our problem as a classical mechanics counterpart of a linear triatomic molecule.

  We analyze  the dynamics arising at total energies $h$ equal to and above the dissociation level $h=0$. Specifically, we investigate escape or scattering solutions, with focus on \textit{$2+1$ escape configurations} (or binary formation), where  two atoms remain close while the third separates,   and  \textit{$1+1+1$ escape configurations}, where all three inter-particle separations infinitely increase as $t \to \infty.$  

 Our approach relies on the choice of coordinates. Specifically,   we use McGehee-type  coordinates, where the size of the system, thought of as the radius  of a sphere in the norm given by the moment of inertia,  becomes one of the variables. Further, using  inversion diffeomorphic  transformations and regularizing the equations of motion, we convert states of infinite size  to states confined to  so-called \textit{infinity manifold}, that is, a compact  manifold glued to the energy level set. 
 In particular, the  $2+1$ and  $1+1+1$ escape configurations appear as invariant submanifolds 
 (lines or equilibria) of the infinity manifold.
  While fictitious, the infinity manifold gives useful information about the real flow via the continuity  of solutions with respect to initial data. 

  Consequently, we analyze the flow on the infinity manifold in both $h=0$ and $h>0$ cases. We prove that on the  zero energy level set, a  zero Lebesque measure set of initial conditions leads to $1+1+1$ escape configurations. For \textit{mass-symmetric} systems, where the outer atoms of the same kind,   we demonstrate the existence  a heteroclinic orbit  connecting two $1+1+1$   asymptotic states that is persistent  for small mass perturbations.  
  For strictly positive energies we prove that the  set of initial conditions leading to  $1+1+1$ escape configurations is of positive Lebesque measure.

 Our  investigation is complemented by  numerical explorations performed for the case $h=0$. These suggest that for most of the initial conditions,  the orbits lead to a $2+1$ escape configurations. For mass-symmetric systems we signal the existence of a periodic orbit that is reminiscent of the Schubart orbit in the collinear three body problem of celestial mechanics (see \cite{Mikkola1}, \cite{Mikkola2} and \cite{Schubart}).  This is in a sense  no surprise, since the Schubart orbit, a periodic solution with two collisions per period, becomes evident  after  the double collisions are regularized, taking the form of elastic bounces. We suspect that a ``Schubart" periodic solution is present in any collinear three mass point system where the pairwise interaction is given by either an attractive  potential regularizable at double collisions, or a small-range-repulsive/large-range-attractive potential. We also believe that the existence of such periodic orbits might be proven using variational methods as in \cite{Venturelli}. We find these interesting questions that we intend to investigate  elsewhere. 

The paper is organized as follows: in Section 2 we review briefly the equations of motion and integrals for a collinear three mass point systems.  We also classify the  escape configurations and find an  invariant  subspace in the case of mass-symmetric systems. In the next section, we introduce McGehee coordinates and  observe that for $h\geq0$ the flow does not admit any equilibrum states. In Section 4,  escape dynamics  is discussed for the case  of zero total energy. We start by converting infinite size states into states at the origin. We further  regularize double collisions, that, while physically impossible, appear as singularities of the vector field. The infinity manifold is defined, which, due to the regularizations of double collisions, shows up  as a compact manifold. The flow on the resulting infinity manifold, its equilibria and invariant manifolds are retrieved. This allows to  determine the Lebesque measure of  the set of initial conditions leading to $1+1+1$ escape configurations.  We conclude the section by presenting some numerically obtained results, including the periodic orbit mentioned above.  In Sections  5, we focus on the case $h>0.$ After applying an appropriate change of space and time coordinates, the infinity manifold $N_h$ appears as a manifold with boundary, namely  a cylinder with caps given by the (fictitious) double collisions. The equilibria  of $N_h$ are classified into six manifolds with boundary.
The dimension of the  stable and unstable manifolds of the equilibria corresponding to $1+1+1$ escape configurations  are then calculated 
using the concepts of overflowing and inflowing invariant manifolds. The main result is stated and proven  in Theorem 5.5 together with Corollary 5.6.

\section{Equations of motion}

Consider a system formed by  three material points (atoms)  of masses $m_1,$ $m_2$ and $m_3,$  constrained  to move on a line, with the pairwise interaction is given by a   potential is of the form (\ref{LJ}). Let the  configuration of the system be described by ${\bf q}=(q_1, q_2, q_3) \in \mathbb{R}^3$, where without loosing generality, we assume an  ordering   $q_1< q_2< q_3$ at all times, and denote  the momenta 
by ${\bf p}:= (p_1, p_2, p_3) \in \mathbb{R}^3.$
The dynamics  is then given  by a canonical Hamiltonian system with the Hamiltonian:
\[H({\bf q}, {\bf p}) = \frac{1}{2}{\bf p}^T M^{-1}{\bf p} + W({\bf q}),\] where $M$ is the mass matrix defined by:
\[M:= \left[ \textrm{diag} (m_i) \right].\] 
 The  potential $W({\bf q})$ consists of  a two homogeneous terms sum:
\[W({\bf q}) = -U({\bf q}) +V({\bf q}), \]
\[U({\bf q}):= \sum \limits_{1 \leq i<j\leq3}\frac{\alpha_{ij} }{|q_i-q_j|^{a}} ,\qquad
V({\bf q}):=   \sum \limits_{1 \leq
i<j\leq3}\frac{ \beta_{ij}}{|q_i-q_j|^{b}}\] with 
$\alpha_{ij}$ and $\beta_{ij}$  real positive parameters and the exponents $a$ and  $b$  such that $4<a<b-1$. In the case of the Lennard-Jones potential,  when $a=6$ and $b=12,$ we have:
\[\alpha_{ij}:= 4\epsilon_{ij}(r^{ij}_{0})^{a}>0, \qquad \beta_{ij}:=\alpha_{ij} (r^{ij}_{0})^{b-a}>0, \] 
where  $\epsilon_{ij}$ and $r^{ij}_{0}$ are the characteristic parameter and the depth of the potential well, respectively,  corresponding to the  interaction of the atoms $i$ and $j.$

Since the potential has
singularities at  the collision points
\[\Delta:= \{    {\bf q} \in
\mathbb{R}^3 : q_1=q_2, \, q_2=q_3, \, q_1=q_3\},\] 
the associated
Hamiltonian vector field
\begin{align} \label{init}
\dot {\bf q}&:= M^{-1}{\bf p}  \\
 \dot {\bf p}&:= -\nabla W \nonumber
\end{align} 
is well-defined on the set $(\mathbb{R}^3 - \Delta) \times \mathbb{R}^3.$ 
Standard results of the differential equations theory ensure, for given 
initial data $(\q(0), \p(0)) \in (\mathbb{R}^3 - \Delta) \times \mathbb{R}^3,$ the existence and uniqueness of an 
analytic solution $(\q(t), \p(t))$, 
defined on a maximal interval $[0, t^*),$ 
$t^* \leq \infty.$ Analogously, one can work
with intervals of the form $(t^*,0].$ 
In case $t^*$ is finite, the solution is said to 
experience a \textit{singularity}. 

Since the system is conservative, along any solution $({\bf q}(t), {\bf p(t)})$ we have:
\[H({\bf q}(t), {\bf p(t)})= \frac{1}{2}{\bf p}^T(t) M^{-1}{\bf p(t)}  +W\left(\q(t) \right) = const.=:h \]
or, equivalently,
 \begin{align} \label{energ1}
 \frac{1}{2}{\bf p}^T(t) M^{-1}{\bf p(t)}  & -
\frac{1}{|q_1(t)-q_2(t)|^{a}}
  \left[ \alpha_{12} -\frac{\beta_{12}}{|q_1(t)-q_2(t)|^{b-a}}  \right]\\
 & 
 -\frac{1}{|q_2(t)-q_3(t)|^{a}}
  \left[\alpha_{23}- \frac{\beta_{23}}{|q_2(t)-q_3(t)|^{b-a}} \right]  \nonumber\\
 &-\frac{1}{|q_3(t)-q_1(t)|^{a}}
  \left[ \alpha_{13}-\frac{\beta_{13}}{|q_3(t)-q_1(t)|^{b-a}}  \right] =h. \nonumber\end{align}

\medskip

 \begin{remark} The ODE system (\ref{init}) has no singular solutions. 
 
 Indeed, assume that as $t\to t^* < \infty$, potential $W\left(\q(t)\right)\to \infty.$. Since the kinetic term is positive for all $t$, the left hand side of \ref{energ1}) approaches infinity, while  the right hand side is constant. This is a contradiction.

 
 This is in agreement with  physical intuition. The repulsive potential term is dominant at short range interactions and thus collisions, double or triple, are impossible.
 

  \end{remark}

 \medskip
 Using the linear momentum  conservation, we  reduce the dimension of the system. Thus, we restrict the configuration space to the linear subspace
\[{\bf Q}:=  \{    {\bf q} \in \mathbb{R}^3 : m_1q_1+m_2q_2+  m_3q_3=0\}  \] and the momenta to
\[{\bf P}:=  \{    {\bf p}  \in \mathbb{R}^3 : { p}_1+{p}_2+{p}_3=0\}  \]
and so the system (\ref{init}) defines a vectorfield free of  singularities on the four dimensional space $({\bf Q} \setminus \Delta) \times {\bf P}.$

\begin{definition}
A collinear atomic-like three mass point system  where the outer mass points  $m_1$ and $m_3$ are of the same kind, i.e. $m_1=m_3,$  $\alpha_{12} = \alpha_{23} $ and $\beta_{12} = \beta_{23}$ is called 
a \textit{mass-symmetric} system.
\end{definition}

\bigskip
\begin{remark}
  For mass-symmetric systems, the Hamiltonian becomes 
 \begin{align} \label{hamXYX}
H({\bf q}, {\bf p}) &= \frac{1}{2m_1}(p_1^2 + p_3^2)+
\frac{1}{2m_2}p_2^2
- \frac{1}{|q_1-q_2|^{a}}
  \left[\alpha_{12} - \frac{\beta_{12}}{|q_1-q_2|^{b-a}}  \right]  \\
  &
  -\frac{1}{|q_2-q_3|^{a}} \left[ \alpha_{12} - \frac{\beta_{12}}{|q_2-q_3|^{b-a}} \right] -\frac{1}{|q_3-q_1|^{a}} \left[\alpha_{13}- \frac{\beta_{13}}{|q_3-q_1|^{b-a}}  \right]. \nonumber
 \end{align}
In this case, a direct verification shows that the subspace 
\[CC:= \{q_2=p_2=0, q_1+q_3 = 0, p_1+p_3 =0\}\]
is invariant  for (\ref{init}). Physically,  $CC$ consists in motions having  $m_2$ fixed in the centre of mass, while  the outer mass points  move   symmetrically with respect to $m_2$.

\end{remark}

\medskip 
We end this section by introducing two definitions.

\smallskip 
\begin{definition} \label{2+1}
A solution $(\q(t), \p(t))$ of  (\ref{init}) is called  a \textbf{$\bf{2+1}$} escape configuration (or binary formation), if 
either
\[\lim \limits_{t \to \infty}|q_1(t) - q_2(t)| =\infty \quad  \text{and} \quad  \lim \limits_{t \to \infty}|q_2(t) - q_3(t)| < \infty,\]
or
\[\lim \limits_{t \to \infty}|q_1(t) - q_2(t)| < \infty  \quad  \text{and} \quad \lim \limits_{t \to \infty}|q_3(t) - q_2(t)| =\infty.\] 
\end{definition}

\smallskip 
\begin{definition}  \label{1-1-1}
A solution $(\q(t), \p(t))$ of  (\ref{init}) is called  a \textit{$\bf{1+1+1}$ escape configuration}  if 
\[\lim \limits_{t \to \infty}|q_2(t) - q_1(t)| =  \lim \limits_{t \to \infty}|q_3(t) - q_2(t)| =\infty.
\] 

\end{definition}

\section{McGehee coordinates}

We introduce a new  set of  coordinates similar to those defined by McGehee in \cite{mcgehee}:
\begin{align*}r&:= \sqrt{{\bf q}^T M{\bf q} } \\
{\bf s} &= r^{-1}{\bf q} \\
v&= r^{b/2}{\bf p}^T {\bf s}\\
{\bf u}&= r^{b/2}\left({\bf p}- v M {\bf s}\right)
\end{align*}
and obtain the following system:
\begin{align}  \label{system1}
\dot r &:= rv \\
 \dot v&= \frac{b}{2}v^2+ {\bf u}^{T} M^{-1}{\bf u}- a r^{b-a}U({\bf s}) +bV({\bf s}) \nonumber \\
\dot {\bf s} &= M^{-1}{\bf u} \nonumber \\
\dot {\bf u}&= \left( \frac{b}{2}-1 \right)v {\bf u}- ({\bf u}^T M^{-1}{\bf u}) M{\bf s} +r^{b-a}\left( \nabla U({\bf s})  +a U({\bf s}) M{\bf s}  \right) - \left(\nabla V({\bf s})+b V({\bf s})M{\bf s}   \right). \nonumber
\end{align} 
Thus $r^2$ is the moment of inertial of the system and escape  corresponds to $r \to \infty.$ Let
\begin{equation}\label{def:S} {\bf S}= \{{\bf q} \in{\bf Q} \, : \, r^2={\bf q} M {\bf q}=1\}\end{equation}
be the unit sphere in $\bf Q$ in the norm given by the moment of inertia.   A 
point $\s \in {\bf S}$ defines a configuration for the system of particles and $(r, \s) \in (0, \infty) \times {\bf S},$ $(r, \s) \to r\s$ can be thought of as polar coordinates on $\bf Q \setminus \Delta.$  
Since   $\u^T \s=0,$ the velocity is broken into  a rescaled  radial component $v$ and a rescaled  tangential component ${\bf u},$ 
and the bundle 
\[{\bf T}:= \{(\q, \p)\in ({\bf Q} \setminus \Delta) \times {\bf P} \,| \, \q \in {\bf S}, \p^T \, \q =0\}\]  
can be thought of as the tangent bundle on $\bf S.$ 
Note that radial velocity $v$ measures the rate of change of the size of the system, where as $\u$ defines the rate of change of the configuration.

The map
\begin{align*}
(0, \infty) \times \mathbb{R} \times {\bf T} &\to ({\bf Q} \setminus \Delta) \times {\bf P}\\
(r, v, \s, \u ) & \to (\q, \p):= (r \s, r^{-b/2} \u + v M \s)
\end{align*}
is a real analytic diffeomorphism and in the new $(r, v, \s, \u)$ coordinates  the energy integral becomes:
\[\frac{1}{2} \left( {\bf u}^T M^{-1}{\bf u} +v^2  \right) -r^{b-a}U({\bf s} ) +V({\bf s} )=hr^{b}. \]

\medskip
We further reduce the dimension of the system following ideas introduced by  McGehee (\cite{mcgehee}). For reader's convenience, we repeat here part of McGehee's construction. For more details, see  \cite{mcgehee}. 
We assume, without loosing any generality,  that the particles are ordered on the line such that $q_1 <q_2<q_3.$ The ordering
is preserved along any orbit. Let
\begin{align*}
{\bf S}_0 =\{{\bf s} \in {\bf S}: s_1 <s_2<s_3 \}, \quad
&{\bf T}_0 = \{({\bf s} , u) \in {\bf T}: {\bf s}  \in {\bf S}_0 \}, \\
{\bf S}_1=\{{\bf s} \in {\bf S}: s_1 \leq s_2 \leq s_3\},  \quad
&{\bf T}_1 =\{( {\bf s} , u) \in {\bf T}: {\bf s}  \in {\bf S}_1\}.
\end{align*}
and let ${\bf a}=(a_1,a_2,a_3)$ and ${\bf b}=(b_1,b_2,b_3)$ be
the unique points on {\bf S} with $a_1 =a_2<a_3$ and $b_1
<b_2=b_3$. Thus, the set ${\bf S} _1$ is a closed interval 
 with endpoints ${\bf a}$ and ${\bf b}$, while ${\bf S}_0$  is the corresponding open interval.
The endpoints ${\bf a}$ and ${\bf b}$  correspond to double collisions and  are given by:
\[a_1=a_2=-\sqrt{\frac{m_3}{(m_1+m_2)(m_1+m_2+m_3)}}<a_3 =\sqrt{\frac{m_1+m_2}{m_3(m_1+m_2+m_3)}} \,,\]
\[b_1=-\sqrt{\frac{m_2+m_3}{m_1(m_1+m_2+m_3)}}<b_2=b_3= \sqrt{\frac{m_1}{(m_2+m_3)(m_1+m_2+m_3)}}\,.\]

\medskip
\noindent
 Next we define a diffeomorphism between $[-1,1] \times \mathbb{R} $ and   ${\bf T}_1$. Let
\[A_1:=\left[ \begin{array}{ccc}
1\,&\,\,1\,\,&1 \\
1\,&\,\,1\,\,&1 \\
1\,&\,\,1\,\,&1
\end{array}
\right]  \qquad
A_2:=\left[ \begin{array}{ccc}
\,\,0 & \,\,\,\,\,1 & \,-1\\
-1  & \,\,\,\,\,0 & \,\,\,1 \\
\,\,1 &\,-1 & \,\,\,0
\end{array}
\right]
\]
and
\[
\tilde A:=\left(\frac{1}{m_1+m_2+m_3} \right)A_1M + \left( \frac{m_1\,m_2
\,m_3}{m_1+m_2+m_3}\right)^{1/2}M^{-1}A_2.
\]
Then we have $\tilde  A^T M \tilde  A =M$,   $\tilde  A {\bf q } \in {\bf Q}$  for all ${\bf q } \in {\bf Q},$ and
\begin{align} \label{prod-zero}
    &{\bf q }^T M  \tilde A \,{\bf q }=0,    \\
    &\tilde  A^2 {\bf q }=-{\bf q }.   \label{prod-zero-bis}
\end{align}
Also,   ${\bf a }^T \tilde  A^T M \,{\bf b }>0.$ So, considering  {\bf Q}
with the inner product induced by M, the matrix $\tilde  A$ is a rotation by
$\pi/2$ in ${\bf Q}$, ${\bf a}$ and ${\bf b}$ have unit length,
and $\{{\bf a},\tilde   A{\bf b} \}$ is an orthonormal basis for ${\bf
Q}$.
Note that ${\bf a}^T M\,{\bf b}$ is a constant depending only on the masses and that $0<{\bf a}^T M\, {\bf b}< 1.$ Let $\lambda$  be the smallest positive number such that
\[\cos {2 \lambda} = {\bf a}^T M\, {\bf b}\] and note that $\lambda$ is a {\textit{constant depending  on the masses only}} and $0< \lambda < \pi/4.$
The map $S: [-1,1] \to {\bf S}_1$ defined by:
\begin{align}\label{define-S} 
s \to {\bf s}= S(s):=(\sin 2\lambda)^{-1} \left [(\sin \lambda(1-s)) {\bf a}+(\sin \lambda(1 +s)){\bf b} \right]
\end{align}
is a real analytic diffeomorphism and the following relation is true (see \cite{mcgehee} for proof):
\begin{equation}\label{S-prim}
S'(s) = \lambda \tilde  A S (s).
\end{equation}
Let
$s:=S^{-1}({\bf s})$ and $u: ={\bf s}^T \tilde  A^T\,{\bf u}.$ Then $s \in[-1, 1]$ and $u\in \mathbb{R}$ and we have  a real analytic diffeomorphism
\begin{align*}
 [0, \infty) \times \mathbb{R}  \times  [- 1, 1] \times \mathbb{R}  &\to
 [0, \infty) \times \mathbb{R}  \times  {\bf T}_1  \\
(r,v,s,u) &\to \left(r,v, S(s), u M \tilde  A S(s) \right).
\end{align*}
which, when restricted to $[0, \infty) \times \mathbb{R}  \times  (- 1, 1) \times \mathbb{R} ,$ is a diffeomorphism onto $[0, \infty) \times \mathbb{R}  \times  {\bf T}_0.$
Finally,   the potential terms are transformed by the maps:
\begin{align*}
U: (-1, 1) \to \mathbb{R}\,, \,\,\,\,s &\to \tilde U(s) :=U(S(s))\\
V: (-1, 1) \to \mathbb{R} \,, \,\, \,\,s &\to \tilde V(s) :=V(S(s)).
\end{align*}
and are given by:
\begin{align} \label{UdeS}
\tilde U(s) &= \sin^a 2 \lambda \left[
\frac{\alpha_{12}}{ \left[ (b_2-b_1) \frac{\sin \lambda (1+s)}{(1+s)} \right]^a (1+s)^a }
+  \frac{\alpha_{23}}{ \left[ (a_3-a_2) \frac{\sin \lambda (1-s)}{(1-s)} \right]^a (1-s)^a }  \right.  \nonumber \\
& \left. +   \frac{\alpha_{13}} {   \left(  \left[    (b_2-b_1) \frac{\sin \lambda (1+s)}{(1+s)} \right] (1+s) + \left[ (a_3-a_2) \frac{\sin \lambda (1-s)}{(1-s)} \right] (1-s)  \right)^a }  \right],
\end{align}
\begin{align} \label{VdeS}
\tilde V(s) &= \sin^b 2 \lambda \left[
\frac{\beta_{12}}{ \left[ (b_2-b_1) \frac{\sin \lambda (1+s)}{(1+s)} \right]^b (1+s)^b }
+  \frac{\beta_{23}}{ \left[ (a_3-a_2) \frac{\sin \lambda (1-s)}{(1-s)} \right]^b (1-s)^b }  \right.  \nonumber \\
& \left. +   \frac{\beta_{13}} {   \left(  \left[    (b_2-b_1) \frac{\sin \lambda (1+s)}{(1+s)} \right] (1+s) + \left[ (a_3-a_2) \frac{\sin \lambda (1-s)}{(1-s)} \right] (1-s)  \right)^b }  \right].
\end{align}
The potential terms $\tilde U$ and $\tilde V$ are strictly positive  convex functions, with $s=\pm1$ boundary     singularities  of degree $(-a)$ and $(-b),$ respectively. In physical terms, $s=\pm1$ represent   double collisions. 
%
%
For mass-symmetric systems,   both  $\tilde U$ and $\tilde V$ are symmetric under the transformation $s \to (-s),$ with $s_0=0$  a common critical point.

 In the new coordinates, system (\ref{system1}) becomes
\begin{align}  \label{system2}
\dot r &:= rv \\
\dot v&= \frac{b}{2}v^2+ u^2- a r^{b-a} \tilde U({ s}) +b \tilde V({ s}) \nonumber \\
\dot {s} &= \frac{1}{\lambda}u \nonumber \\
\dot { u}&= \left( \frac{b}{2}-1 \right)vu +r^{b-a} \frac{1}{\lambda} \frac{d \tilde U(s)}{ds}  -\frac{1}{\lambda} \frac{d \tilde V(s)}{ds}    \nonumber
\end{align}
and the energy conservation reads:
\[\frac{1}{2} \left(u^2 +v^2  \right) -r^{b-a} \tilde U(s) + \tilde V(s)=hr^{b}. \]
Since we considered  $a<b-1,$ the above vectorfield is of differentiable class.

\medskip
\begin{remark} There are no   equilibria for positive energy levels. 

Indeed, equating the right hand side of system (\ref{system2}) to zero, and taking into account the conservation of energy, the following relations must be satisfied:
\begin{align*}
&u=v=0\\
&(b-a)\tilde U(s)=(-h)b r^a\\
& r^{b-a}  \frac{d \tilde U(s)}{ds} - \frac{d \tilde V(s)}{ds}=0.
\end{align*}
Since $\tilde U(s)>0$, it follows that for  $h\geq 0,$ the second equation above cannot be satisfied.
\end{remark}

\section{Dynamics for zero energy}

\subsection{Regularization of the motion}

The main focus of this section is on unbounded motions on the zero energy level.
%
We begin by applying the change of coordinates
$\rho =  {1}/{r}$
together with 
\[x:=u \rho ^{(b-a)/2} \,, \quad y:= v \rho ^{(b-a)/2 }.
\]
Introducing ${d \tau}/{d t} = \rho ^{-(b-a)/2}\,$ as a new time parametrization, 
system (\ref{system2}) becomes:
\begin{align} \label{syst-rho1}
\dot \rho &= -\rho y \\
\dot y &= \frac{a}{2}y^2 +x^2 - a \tilde U(s) +b \rho^{(b-a)}  \tilde V(s) \nonumber \\
\dot s&= \frac{1}{\lambda} x \nonumber \\
\dot x&= \left( \frac{a}{2}-1 \right) xy + \frac{1}{\lambda} \tilde U'(s) -\frac{1}{\lambda} \rho^{b-a}   \tilde V'(s) ,\nonumber
\end{align} 
where we keep the same ``dot" notation for the derivative $d/d\tau$, and where ``prime" means derivation with respect to the $s$ variable.
The energy integral reads:
\begin{equation} \label{energy-rho11}
\frac{1}{2}(x^2+y^2)=  \tilde U(s)-\rho^{b-a} \tilde V(s).
\end{equation}
The vectorfield (\ref{syst-rho1})  is well-defined for $(\rho, y, s, x) \in [0,\infty) \times \mathbb{R} \times (-1,1)\times \mathbb{R}.$   Escape orbits  correspond now to orbits where  $\displaystyle{\lim\limits_{\tau \to \infty} \rho(\tau) =0 }.$ In this case, the \textit{asymptotic escape configuration} is given by $\displaystyle{\lim\limits_{\tau \to \infty} s(\tau)},$ and 
 the  \textit{asymptotic velocity of the system's size} by   $\displaystyle{\lim\limits_{\tau \to \infty} y(\tau)}.$

\medskip
 From the energy relation (\ref{energy-rho11}), we deduce that $\rho$ is bounded   by the domain:
 \[\mathcal{D}:=\left\{(\rho, y, s, x) \in (0,\infty) \times \mathbb{R} \times (-1,1)\times \mathbb{R} : \rho^{b-a} \leq  \frac{ \tilde U(s) } { \tilde V(s) }\right\} \,.\]
 Since both  $ \tilde U(s)$ and $  \tilde V(s)$ are strictly positive, $\mathcal{D}$ is non-empty for any choice of the parameters.

\medskip

  We now extend the vectorfield to  include the double-collision endpoints
  $s = \pm 1$   by using a rescaling of the radial coordinates $\rho$  together with  a Sundman type regularization.
  More precisely, we use two appropriate positive functions $\phi(s)$ and $\beta(s)$  such that by applying the change of variables $(\rho, y,s,x) \rightarrow (R, y,s,w)$   where
 \begin{equation} \label{c-v-1}
  \rho:= \phi(s)R, \quad x:=\beta (s)w
  \end{equation}
  together with an appropriate time reparametrization,
\begin{equation} \label{c-v-2}
  \frac{d \tau}{ d \sigma} = \frac{1}{\beta^2(s)}\, ,
  \end{equation}
  the vectorfield  becomes  well-defined for all $(R,y,s, w ) \in [0, \infty) \times \mathbb{R} \times [-1,1] \times \mathbb{R}$.  
%
 %
 %
 An  optimal choice for $\phi$ and $\beta$ is given by:
  \begin{equation} \label{c-v}
 \phi(s)= 1-s^2, \quad \beta (s)= \frac{1}{(1-s^2)^{a/2}},
  \end{equation} thus obtaining:
   \begin{align} \label{syst-Rh=0-old}
\dot R &= - R\left[   (1-s^2)^a  y - \frac{1}{ \lambda} 2s(1-s^2)^{\frac{a}{2}-1}w  \right]  \\
\dot y &= \left(\frac{a}{2} -1 \right) (1-s^2)^a\left( y^2 - 2 \tilde U(s)\right)  + b   R^{b-a} (1-s^2)^b \, \tilde V(s)+w^2 \nonumber \\
\dot s&= \frac{1}{\lambda} (1-s^2)^{\frac{a}{2}} \, w \nonumber \\
\dot w &= \left( \frac{a}{2}-1 \right) (1-s^2)y w -
\frac{1}{\lambda} a s(1-s^2)^{\frac{a}{2}-1}\, w^2 \nonumber
\\ & \quad  \quad   \quad  \quad  \quad +\frac{1}{\lambda}
(1-s^2)^{\frac{3a}{2}} \,   \tilde U'(s) -\frac{1}{\lambda}
R^{b-a} (1-s^2)^{b+\frac{a}{2}} \, \tilde V'(s). \nonumber
\end{align}
 with the energy integral:
  \begin{equation} \label{energy-Rh=0}
  \frac{1}{2}w^2 + \frac{1}{2} (1-s^2)^{a}\left (y^2  -2 \tilde U(s)\right)+ R^{b-a} (1-s^2)^b \tilde V(s) =0.
  \end{equation}
Thus, at the double-collision endpoints $s=\pm1$ now we have
\begin{align*}
\lim \limits_{s\to (-1)^+} (1-s^2)^a\tilde U(s)=\left(\frac{2 \sin 2\lambda}{\lambda}\right)^a \frac{\alpha_{12}}{(b_2-b_1)^a} =:K_- <\infty \\
\lim \limits_{s\to 1^-} (1-s^2)^a\tilde U(s)=\left(\frac{2 \sin 2\lambda}{\lambda}\right)^a \frac{\alpha_{23}}{(a_3-a_2)^a} 
=:K_+<\infty\\
 \lim \limits_{s\to (-1)^+} (1-s^2)^b\tilde V(s) = \left(\frac{2 \sin 2\lambda}{\lambda}\right)^b \frac{\beta_{12}}{(b_2-b_1)^b} 
=:L_-<\infty\\
 \lim \limits_{s\to 1^-} (1-s^2)^b\tilde V(s) = \left(\frac{2 \sin 2\lambda}{\lambda}\right)^b \frac{\beta_{23}}{(a_3-a_2)^b}
 =:L_+<\infty
\end{align*}
and
\begin{align*}
\lim\limits_{{s\to(-1)^+}}
 \left(\frac{d}{ds}  (1-s^2)^a\tilde U(s)\right) 
  = - \frac{a}{2}\, K_-  \quad \,\,\,
   \lim\limits_{{s\to1^-}}
 \left(\frac{d}{ds}  (1-s^2)^a\tilde U(s)\right) 
  = \frac{a}{2}\, K_+\\
\lim\limits_{{s\to(-1)^+}}
 \left(\frac{d}{ds}  (1-s^2)^b\tilde V(s)\right) 
  = - \frac{b}{2}\, L_-  \quad \,\,\,
   \lim\limits_{{s\to1^-}}
 \left(\frac{d}{ds}  (1-s^2)^b\tilde V(s)\right) 
  = \frac{b}{2}\, L_+
\end{align*}
It follows that system (\ref{syst-Rh=0-old}) it is at least  differentiable for all $(R,y,s, w ) \in [0, \infty) \times \mathbb{R} \times [-1,1] \times \mathbb{R}$  provided $4<a< b-1.$  The domain of motion for $R$ is 
    \[\mathcal{D}= \left\{(R, y, s, w)\in [0, \infty) \times \mathbb{R} \times [-1,1] \times \mathbb{R} \quad :  R^{b-a} \leq    \frac{(1-s^2)^{a} \tilde U(s) }{ (1-s^2)^{b} \tilde V(s)} \right\}.
  \]
Since both $(1-s^2)^{a} \tilde U(s)$ and $(1-s^2)^{b} \tilde V(s) $ are well-defined and continuous on $[-1,1],$  it follows that they attained their extrema.  Thus $R$ is finite at all times.

\medskip
\begin{remark} \label{diss-def-change}
Under the above reparametrizations,  escape definitions (\ref{2+1})  and (\ref{1-1-1}) read as follows:

\medskip
\noindent
A $2+1$ \textrm{escape configuration}, (or \textrm{binary formation}), is a solution of (\ref{syst-Rh=0-old}) such that $\lim \limits _{\sigma \to \infty} R(\sigma) =0$ and $\lim \limits _{\sigma \to \infty} s(\sigma) =\pm 1.$  

\medskip
\noindent
 A $1+1+1$ \textrm{escape configuration} is a solution of (\ref{syst-Rh=0-old}) such that   $\lim \limits _{\sigma \to \infty} R(\sigma) =0$ and $\lim \limits _{\sigma \to \infty} |s(\sigma)|<1.$
\end{remark}

\medskip
\begin{remark} 
For any escape solution, the asymptotic configuration and the  asymptotic  velocity  of the system's size are given by 
$\lim\limits _{\sigma \to \infty} s(\sigma)$ and $\lim\limits _{\sigma \to \infty} y(\sigma),$ respectively. 

\end{remark}

\subsection{Dynamics for zero energy}

From  the energy integral we have:
  \begin{equation} \label{energy-Rh=01}
R^{b-a} (1-s^2)^b \tilde V(s)=(1-s^2)^{a}\left (\tilde U(s) -\frac{1}{2}y^2\right) -  \frac{1}{2}w^2 \geq0
  \end{equation}
which  substituted into the second and the last  equation of (\ref{syst-Rh=0-old}) leads to:
   \begin{align} \label{syst-Rh=0}
\dot R &= - R\left[   (1-s^2)^a  y - \frac{1}{ \lambda} 2s(1-s^2)^{\frac{a}{2}-1}w  \right]  \\
\dot y &= -\left(\frac{b}{2} -1 \right) w^2 +\left(b-a-2\right)   (1-s^2)^a 
\left (\tilde U(s) -\frac{1}{2}y^2\right) 
\nonumber \\
\dot s&= \frac{1}{\lambda} (1-s^2)^{\frac{a}{2}} \, w \nonumber \\
\dot w &= \left( \frac{a}{2}-1 \right) (1-s^2)y w +
\frac{1}{2\lambda} (1-s^2)^{\frac{a}{2}-1}\left[ \frac{(1-s^2)\tilde V'(s)}{\tilde V(s)}-2as  \right] w^2 \nonumber
\\ & \quad  \quad   \quad  \quad  \quad +\frac{1}{\lambda}
(1-s^2)^{\frac{3a}{2}} \,  \left[ \tilde U'(s) -\frac{\tilde V'(s)}{\tilde V(s)}\left (\tilde U(s) -\frac{1}{2}y^2\right)  \right].
\nonumber\end{align}
The dynamics  decouples since  the last three equations may be solved independently  (if the functions $\left(y(\cdot),s(\cdot),w(\cdot)\right)$ are determined, then $R(\cdot)$ is given by relation (\ref{energy-Rh=01})). We thus focus on the  reduced system: 
  \begin{align} 
  \label{red-syst-Rh=0}
\dot y &= -\left(\frac{b}{2} -1 \right) w^2 +(b-a-2)   (1-s^2)^a 
\left (\tilde U(s) -\frac{1}{2}y^2\right) 
\nonumber \\
\dot s&= \frac{1}{\lambda} (1-s^2)^{\frac{a}{2}} \, w\\
\dot w &= \left( \frac{a}{2}-1 \right) (1-s^2)y w +
\frac{1}{2\lambda} (1-s^2)^{\frac{a}{2}-1}\left[ \frac{(1-s^2)\tilde V'(s)}{\tilde V(s)}-2as  \right] w^2 \nonumber
\\ & \quad  \quad   \quad  \quad  \quad +\frac{1}{\lambda}
(1-s^2)^{\frac{3a}{2}} \,  \left[ \tilde U'(s) -\frac{\tilde V'(s)}{\tilde V(s)}\left (\tilde U(s) -\frac{1}{2}y^2\right)  \right],
\nonumber
\end{align} 
where $(y,s, w ) \in  \mathbb{R} \times [-1,1] \times \mathbb{R}$ such that

\[ (1-s^2)^{a}\left(\tilde U(s) -\frac{1}{2} y^2\right) - \frac{1}{2}w^2
\geq0.\]

\begin{remark}
\label{symm1}
The  flow of the system (\ref{red-syst-Rh=0})  is symmetric under  $(y,s,w, \sigma) \to (-y, s, -w, -\sigma).$ 
\end{remark}

\bigskip
The equilibria  are found   at the points $P^{\pm}:=( \pm \sqrt{2 \tilde U(s_{e})}, s_e, 0),$ where $s_e$ is the critical point of $U(s).$ 
From relation (\ref{energy-Rh=01}), we deduce that at the equilibrium $R=0$ and thus $P^{\pm}$ correspond to escape orbits. Also, since $|s_e| \neq 1,$  the equilibria $P^{\pm}$ correspond to $1+1+1$ escape configurations. In particular, we have:

\medskip
\begin{remark}\label{asymp-equil}
For zero energy, in the case of  $1+1+1$ escape configurations, the asymptotic velocity of the system's size  is finite.
\end{remark}

\medskip
  The local behaviour near equilibria is determined using the  linearization of (\ref{red-syst-Rh=0}) at $P^{\pm}.$
 A direct calculation shows that at $P^+$ the eigenvalues are given by
\[ \xi_1=-(1-s_e^2)^a \,  \sqrt{2 \tilde U(s_e)} <0\] and $\xi_{2}<0$ and $\xi_{3}>0,$ the roots of
\begin{equation}
\label{pol}
P(\xi) =
 \xi^2 +
 \left(\frac{a}{2}-1  \right) (1-s_e^2)^a \, \sqrt{2 \tilde U(s_e) } \, \xi
 - \frac{ (1-s_e^2)^{2a}}  {\lambda^2}  \frac{d^2\tilde U}{ds^2}\Big|_{s_e} =0.
\end{equation}
Note that the eigenvector $v_{\xi_1}$ corresponding to $ \xi_1$ is aligned to the $y$ axis, where as the eigenvectors $v_{\xi_2}$ and $v_{\xi_3}$ are directions in the  $s-w$ plane tangent to $M$ at $P^+.$

  Since
all eigenvalues have non-zero real part, the rest point at  $P^+$ is hyperbolic. We deduce that  this equilibrium has a two dimensional stable
manifold and a one dimensional unstable manifold.  An analogous
calculation shows that the stable manifold of the equilibrium at $P^{-}$ is one
dimensional, whereas the unstable manifold is two dimensional.
So we have proved:

\medskip
\begin{proposition}\label{dim-equil}
On  the  zero energy level,    the flow of the collinear attractive-repulsive 3-particle problem, as defined by system (\ref{red-syst-Rh=0}),  has two hyperbolic rest points,  denoted $P^{\pm}.$ Their stable manifold $W^s$ and unstable manifold $W^u$ have dimensions as follows:
\end{proposition}
\begin{align*} \textrm{dim} \, W^u(P^+)=1 \qquad \textrm{dim} \, W^s(P^+)=2 \\
 \textrm{dim} \, W^u(P^-)=2 \qquad \textrm{dim} \, W^s(P^-)=1.
\end{align*}

\medskip

Since the equilibria at $P^{\pm}$ correspond to $1+1+1$ dissociations, we have the following:
\medskip
\begin{corollary} \label{Lebesgue}
In the linear attractive-repulsive 3-particle problem  given by (\ref{init}) with zero energy, the set of initial condition leading to  $1+1+1$ escape configurations has  zero Lebesgue measure.
\end{corollary}


\bigskip

The dynamics of  (\ref{red-syst-Rh=0}) admits the manifold
\begin{equation}
M:=\{(y,s,w)\in  \mathbb{R} \times [-1,1] \times \mathbb{R}  \,: \, (1-s^2)^{a}\left ( \tilde U(s) - \frac{1}{2}y^2  \right)-\frac{1}{2}w^2=0\},
\end{equation} 
 invariant.
 We call  $M$ the \textit{infinity manifold}; it is a  two dimensional boundary submanifold  glued to the three-dimensional zero energy surface and it  is formed  by orbits with $R=0$ for all times. Thus, orbits on
 $M$  represent  fictitious motions where size of the
system is infinite. The  real flow is located in the region bounded by $M$ containing the origin. By
using the  good differentiable properties at the boundaries and the continuity with respect to initial data, we can extract information about  real orbits  close to   those on $M.$

The dynamics on $M$ is given by:
 \begin{align} \label{inf-syst-h=0}
\dot y &= -\left(\frac{a}{2} -1 \right) w^2
 \nonumber \\
\dot s&= \frac{1}{\lambda} (1-s^2)^{\frac{a}{2}} \, w \nonumber \\
\dot w &= \left( \frac{a}{2}-1 \right) (1-s^2)y w -\frac{1}{\lambda}
as(1-s^2)^{\frac{a}{2}-1}w^2
 \nonumber \\ & \quad  \quad   \quad  \quad  \quad
+\frac{1}{\lambda}  (1-s^2)^{\frac{3a}{2}} \,   \tilde U'(s). \nonumber
\end{align}    
It is immediate that the flow  is decreasing along the $y$ coordinate, since $\dot y$ is always negative.   The boundaries $\{s=\pm 1\}$ appear as invariant manifolds. They are four  one dimensional boundary submanifolds glued to $M,$ given by
     \begin{equation} \label{DC-man}
DC^{\pm}:= \{(y, s, w) \in  \mathbb{R} \times [-1,1] \times \mathbb{R}\,: \,s =\pm1, \quad  \frac{1}{2}w^2  -K_{\pm} =0\}
\end{equation}
 Since $\{s \pm 1\}$ correspond to fictitious double collision configurations, we call $DC^{\pm}$  {\textit{the double collision manifolds}}.  The dynamics on $DC^{\pm}$  is given by, respectively:
\begin{align*}
\dot y &= -2 \left(\frac{a}{2} -1 \right) K_{\pm}   \\
\dot w &=  0. \nonumber
\end{align*}
Orbits tending to $DC^{\pm}$  correspond to  $2+1$ escape configuration. It follows that in this case, $y,$ the asymptotic  velocity  of the system's size, is unbounded. In other words,    for a binary formation, the system is uniformly increasing in size. This is in contrast to  the case of $1+1+1$ escape configurations, where 
 the asymptotic  velocity of the system's size is finite.

\begin{figure}[tbp]

\begin{center}\includegraphics[angle=0,
scale=0.65]{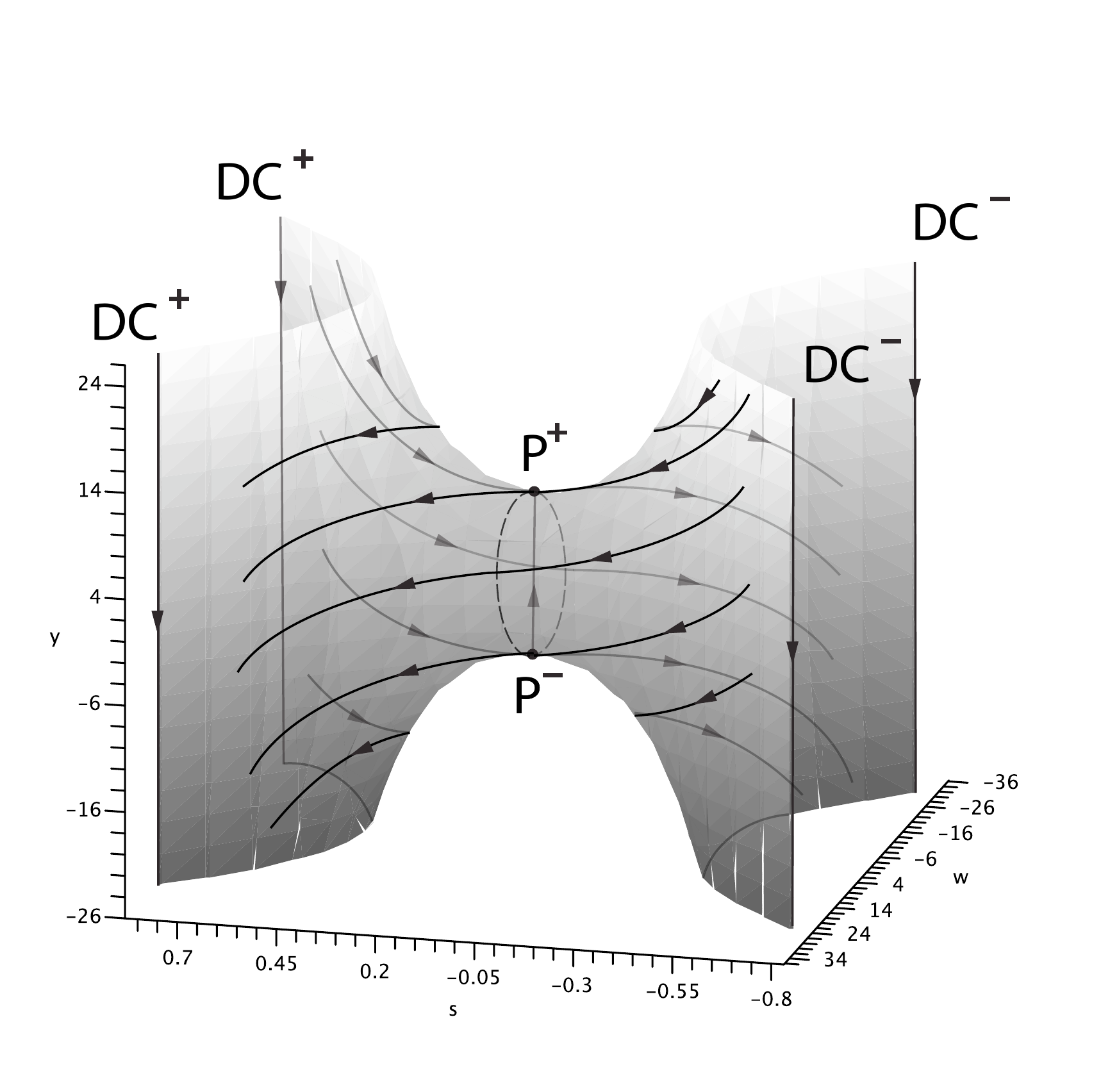} \caption{The zero-energy infinity manifold together with its  flow and the heteroclinic orbit joining the $1+1+1$ escape configurations $P^-$ and $P^+$. \label{fig-inf-man-flow} }
\end{center}
\end{figure}
In Figure \ref{fig-inf-man-flow}  the qualitative features of the flow on $M$ is depicted for mass-symmetric systems.
For all other systems,
 the picture is qualitatively
the same, only the figure's ``neck"  is shifted to the left or
right of $s=0,$ depending on where the critical point $s_e$ of
$\tilde U(s)$ is situated. 
Using the continuity of solutions with respect to initial data we deduce:

\begin{remark}
There exists orbits connecting two different  $2+1$ escape configurations.
\end{remark}

\begin{remark}
There exists orbits connecting  $2+1$  to  $1+1+1$ escape configurations. 
\end{remark}


\medskip
For mass-symmetric  systems,  due to the symmetry of $\tilde U$ and $\tilde V,$ the flow of (\ref{red-syst-Rh=0}) is symmetric under the transformations 
\begin{equation}\label{symm2}
(y,s,w) \to (y, -s, -w).
\end{equation}
In this case, there exits a  heteroclinic  orbit  connecting the equilibria at $P^-$ and $P^+.$ 
This orbit that may be  determined by integrating
\[\dot y= (b-a)\left( \tilde U(0) -\frac{1}{2} y^2 \right).\]

\begin{proposition} For mass-symmetric systems,  the unstable manifold $W^u(P^-)$ and the stable  manifold $W^s(P^+)$ intersect transversally along the heteroclinic orbit that connects $P^-$ to $P^+.$
\end{proposition}
{\it Proof.~}
Consider an orbit $\gamma_1:= \{\left(y(\sigma), s(\sigma), w(\sigma)\right)\,: \, \sigma \in \mathbb{R} \} \in W^u(P^-).$ Using the symmetries given by Remark \ref{symm1} and relation (\ref{symm2}),  we have
\begin{align*} 
 (-\sqrt{2U(0)}, 0, 0) &= \lim\limits_{\sigma \to -\infty} \left(y(\sigma), s(\sigma), w(\sigma)\right)
 \\&= \lim\limits_{\sigma \to -\infty} \left(y(\sigma), -s(\sigma), -w(\sigma)\right)
 \\&=\lim\limits_{\sigma \to -\infty} \left(-y(-\sigma), -s(-\sigma), w(-\sigma)\right)
 \\&=\lim\limits_{\sigma \to +\infty} \left(-y(\sigma), -s(\sigma), w(\sigma)\right)
 \\&=- \lim\limits_{\sigma \to +\infty} \left(y(\sigma), s(\sigma), -w(\sigma)\right)
\end{align*} and so 
\[ 
  \lim\limits_{\sigma \to +\infty} \left(y(\sigma), s(\sigma), -w(\sigma)\right)=(+\sqrt{2U(0)}, 0, 0).\]
Thus,  $\gamma_2:=\{\left(y(\sigma), s(\sigma), -w(\sigma)\right)\,: \,\sigma \in \mathbb{R}\} \in W^s(P^+)$.

\medskip
Assume $W^u(P^-)= W^s(P^+).$ In particular, it follows that  $ \gamma_1 \in  W^s(P^+).$ Further, near $P^+,$ both $\gamma_1$ and $\gamma_2$ must be tangent to  the stable eigenspace of $P^+.$  Thus, near $P^+ $, $\gamma_1$ and $\gamma_2$ may  be expressed as a linear combination of the stable eigenvectors $v_{\xi_1}$ and $v_{\xi_2}.$ 

Now, $v_{\xi_1}$ is aligned to the $y$ coordinate. Thus, close to $P^+,$  the $s-w$ components of $\gamma_1$ and $\gamma_2$, that is  $\left(s(\sigma), w(\sigma) \right)$ and $\left(s(\sigma), -w(\sigma)\right),$ either are both zero, or are parallel to $v_{\xi_2}$. In the first case we retrieve the heteroclinic orbit that connects  $P^-$ to $P^+.$ In the second case, we must have $w(\sigma)=0,$ and,    since no restriction was done over $\gamma_1,$ this leads to the fact that the stable eigenvector $v_{\xi_2}$  must be parallel to the $s$ axis. The last statement can be easily verified that is not true. So we conclude that $W^u(P^-)$ and $W^s(P^+)$ intersect transversally along the heteroclinic orbit that connects  $P^-$ to $P^+.$

\medskip
  \begin{corollary}The heteroclinic orbit $P^- \, P^+$ is stable under parameter perturbation, and in particular, it will be present in the mass-asymmetric case for small  mass perturbations. 
  \end{corollary}

\subsection{Numerical investigations}

\begin{figure}[tbp]
\begin{center}\includegraphics[angle=0,
scale=0.4]{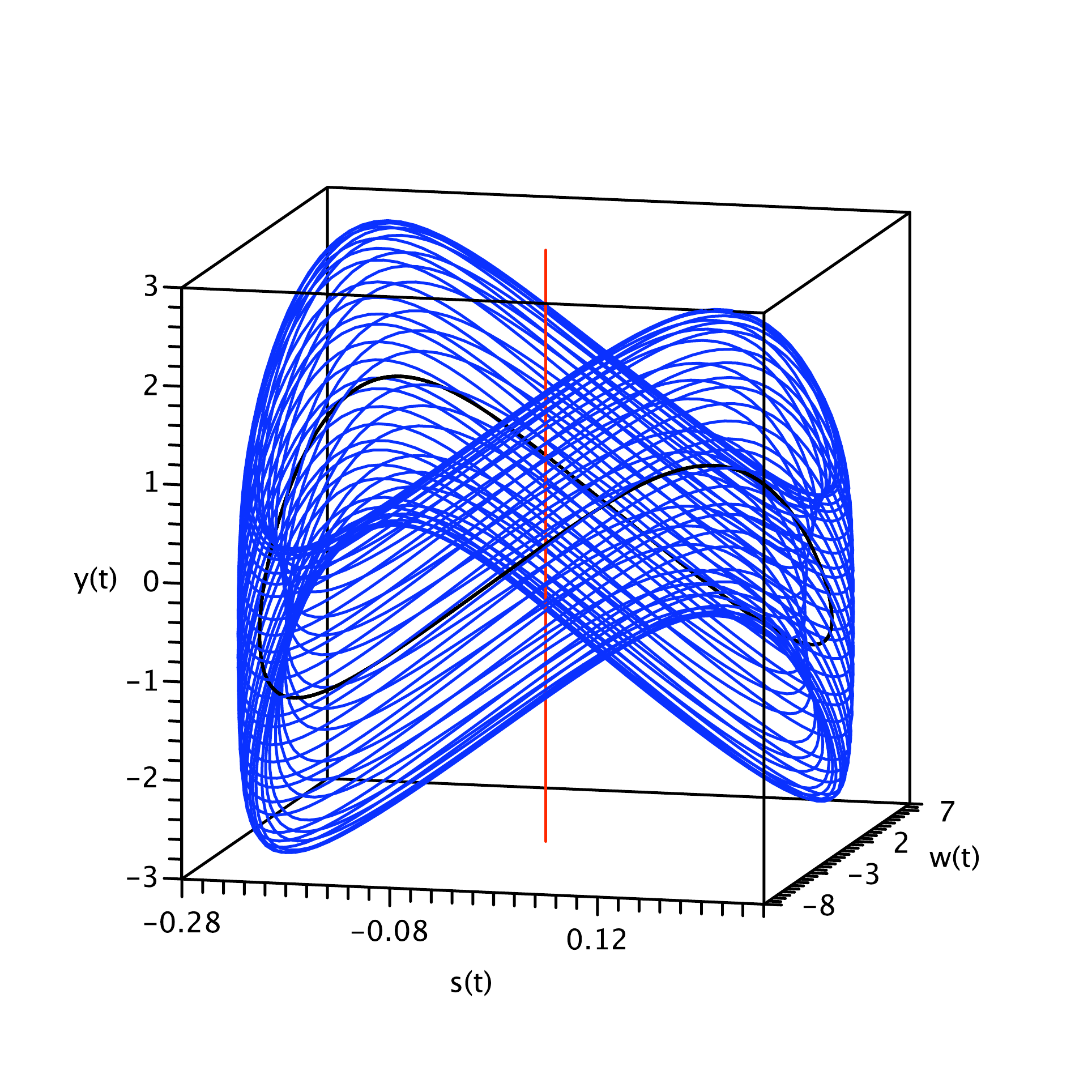} 
\caption{A periodic orbit (PO) for  mass-symmetric molecules together with a  surrounding orbit. The parameters are $m_1=1$ and $m_2=2,$ $\alpha_{12}=1,$ $\beta_{12}=0.5,$ and $\alpha_{13}=2, \beta_{13}=1,$ $a=6,$ $b=12,$ with  initial  conditions    $y(0) = 0, s(0) = 0, w(0) = 6.67$ for the PO, represented by the black curve, and $y(0) = 0, s(0) = 0, w(0) = 6.5$ for the surrounding orbit.   The vertical red segment depicts the heteroclinic orbit $P^- P^+.$
\label{PO} }
\end{center}
\end{figure}

\begin{figure}[tbp]
\begin{center}\includegraphics[angle=0,
scale=0.4]{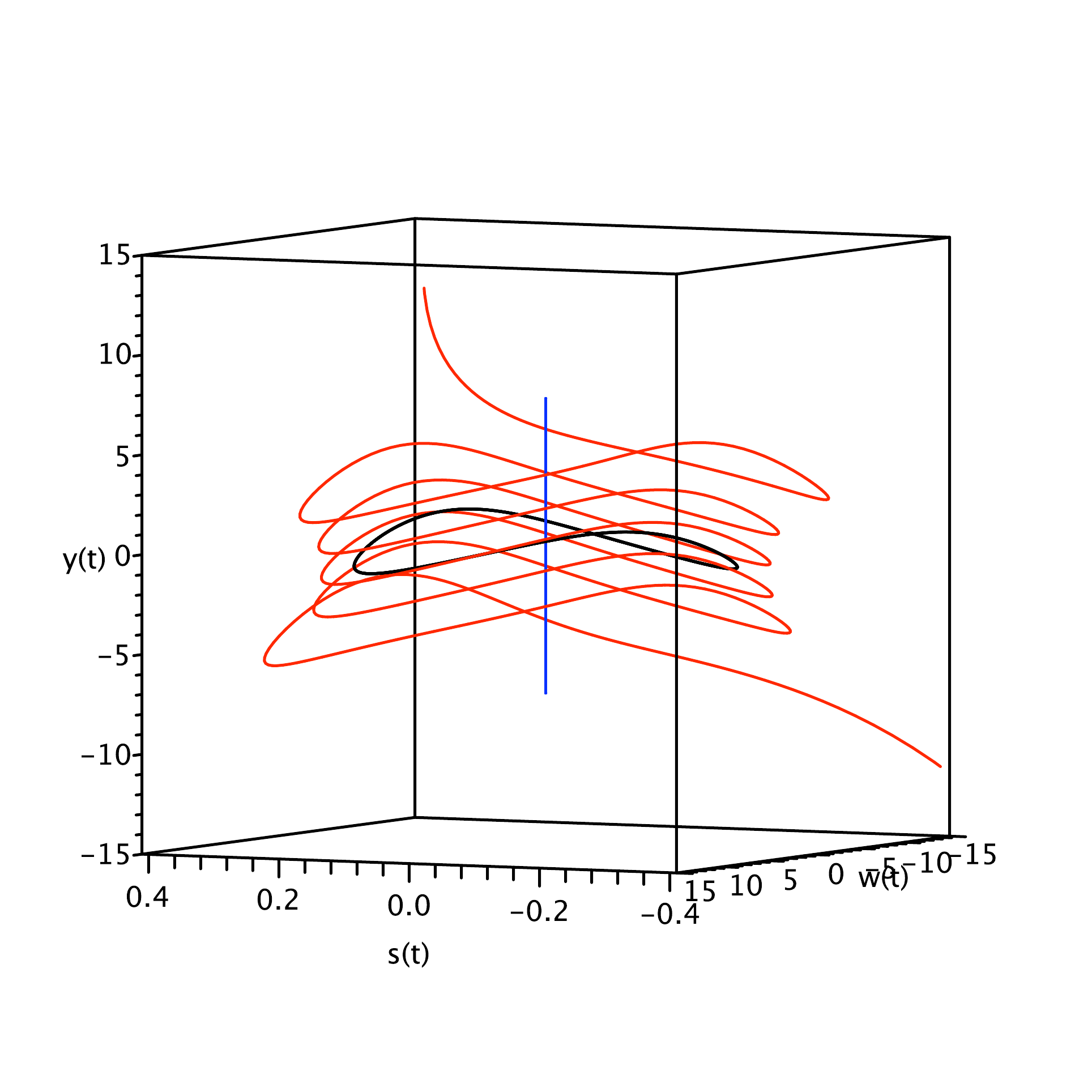} 
\caption{A periodic orbit (PO) for  mass-symmetric molecules together with an  escape orbit. The parameters and the initial  conditions for the PO were taken as  in the  Figure 2. The initial  conditions  for the escape orbit were $y(0) = 0, s(0) = 0, w(0) = 6.9.$
\label{ESPO} }
\end{center}
\end{figure}

We complete our study  by presenting some preliminary numerically explorations. These show that most of the initial conditions lead to  $2+1$ escape configurations. A typical orbit emerges from an asymptotic  $2+1$ configuration, approaches a $1+1+1$ configuration, and falls back into one of the two  possible $2+1$ escape configurations.

 We also noticed that in the mass-symmetric case, a periodic orbit (PO) appears surrounding the heteroclinic orbit $P^- P^+.$ The PO is surrounded by its  normal modes, orbits  that generically   fill in densely a cylindrical band around the PO  - see  Figure \ref{PO} - . The PO seems  stable under perturbations of the parameters such as the ratio of the masses and the Lennard-Jones $\alpha$ and $\beta$ parameters.
 
 As mentioned in the Introduction, the PO is reminiscent of the Schubart orbit discovered in the context of the collinear Newtonian three body problem with regularized  double collisions. We suspect that the Schubart orbit is present  in any collinear mass-symmetric
three body problem  where the pairwise interaction is given by either a 
potential regularizable at double collisions, or a small-range-repulsive/large-range-attractive potential.  We intend to fully investigate this topic elsewhere.

\section{Dynamics for positive energy}

For $h>0$ we consider a different change of variables and rescaling of time
\begin{equation*}
\tilde\rho = 1/r,  \quad \tilde v= \tilde\rho^{b/2}v, \quad \tilde {\bf s}={\bf s}, \quad \tilde {\bf u}= \tilde\rho^{b/2} {\bf u}, \quad d\tau=\tilde \rho^{b/2}d\eta. \\
\end{equation*}
%
%
%
%
System (\ref{system2}) now becomes
\begin{align} 
 \label{systemh>0}
\tilde\rho '&= -\tilde\rho\tilde v \\
\tilde v'&= \tilde{\bf u}^TM^{-1}\tilde{\bf u} -a\tilde\rho^aU(s)+b\tilde\rho^bV(s)  \nonumber \\
\tilde{\bf s}' &= M^{-1}\tilde{\bf u} \nonumber \\
\tilde{\bf u}'&= -\tilde {\bf u}\tilde v-\tilde{\bf u}^TM^{-1}\tilde{\bf u}M\tilde{\bf s}+\tilde\rho^a(\nabla U(\tilde{\bf s})+aU(\tilde{\bf s})M\tilde{\bf s})-\tilde\rho^b(\nabla V(\tilde{\bf s})+aV(\tilde{\bf s})M\tilde{\bf s}). \nonumber
\end{align}
%
%
%
together with the energy relation
\[
\frac 1 2(\tilde{\bf u}^TM^{-1}\tilde{\bf u} +\tilde v^2)-\tilde\rho^aU(\tilde{\bf s})+\tilde \rho^bV(\tilde{\bf s})=h.
\]
As in Section 3, we reduce the dimension of the system  by introducing the new variables
\[
\tilde s=S^{-1}(\tilde {\bf s}), \quad \tilde u=\tilde{\bf s}^TA^T\tilde{\bf u}.
\]
We also use the potentials $\tilde U$ and $\tilde V$ introduced in equation
(\ref{UdeS}) and (\ref{VdeS}).
In these reduced coordinates  system (\ref{systemh>0}) reads
\begin{equation}
\begin{split}  \label{systemh>0-1}
\tilde\rho'&= -\tilde\rho\tilde v \\
\tilde v'&= \tilde u^2 -a\tilde\rho^a\widetilde U(\tilde s)+b\tilde\rho^b \widetilde V(\tilde s)   \\
\tilde s' &= \frac 1 \lambda \tilde u  \\
\tilde u'&= -\tilde u\tilde v+\frac 1 \lambda \tilde\rho^a \frac{d\tilde U}{d\tilde s}-\frac 1 \lambda \tilde\rho^b \frac{d\widetilde V}{d\tilde s}.
\end{split}
\end{equation}
and the energy integral takes the form
\[\frac 1 2(\tilde u^2+\tilde v^2)-\tilde\rho^a\widetilde U+\tilde\rho^b\widetilde V=h.\]
We further regularize double collisions using a Sundman type regularization.
Let
\[
\tilde \rho=(1-s^2)\tilde R=\Theta(s)\tilde R, \quad\quad \frac{d\tau}{d\sigma}=\Theta(\tilde s)
\]
Then the energy relation becomes
\beq\label{energy_h>0}
\frac 1 2(\tilde u^2+\tilde v^2)-\tilde R^a\Theta(\tilde s)^a \tilde U(\tilde s)+\tilde R^b\Theta(\tilde s)^b \tilde V(\tilde s)=h.
\eeq
Using  the energy relation  we can write the equations of motion as
\begin{equation}
\begin{split} \label{systemh>0-2}
\tilde R'=&-\frac 1 \lambda\frac{d\Theta}{ds}\tilde u R-\Theta \tilde R\tilde v\\
\tilde v'=& \Theta(\tilde s)\left[\tilde u^2 -a \tilde R^a\Theta(\tilde s)^a \tilde U(\tilde s)+b\tilde R^b\Theta(\tilde s)^b \tilde V(\tilde s)\right]\\
\tilde s'=&\frac 1 \lambda \Theta(\tilde s)\tilde u\\
\tilde u'=& -\Theta(\tilde s)\tilde u\tilde v+\frac 1 \lambda \left[\tilde R^a\Theta(\tilde s)^{a+1}\frac{d\tilde U}{ds}-\tilde R^b \Theta(\tilde s)^{b+1}\frac{d\tilde V}{d\tilde s}\right]
\end{split}
\end{equation}
where $'$ now denotes differentiation with respect to $\sigma$. These equations define a real analytic vectorfield  on $\mathfrak R :=[0,\infty)\times{\mathbb R}\times [-1,1]\times {\mathbb R}$. For each  $h>0$ let
\begin{equation} \label{Fh}
F_h=\{(\tilde R,\tilde v,\tilde s,\tilde u)\in \mathfrak R: \frac 1 2(\tilde u^2+\tilde v^2)-\tilde R^a\Theta(\tilde s)^a \tilde U(\tilde s)+\tilde R^b\Theta(\tilde s)^b \tilde V(\tilde s)=h
\}.
\end{equation}
Since the gradient of expression (\ref{energy_h>0}) does not vanish on $F_h$, $F_h$ is a three dimensional real analytic submanifold of $\mathfrak R$. $F_h$ can be written as a union of
\begin{equation} \label{Nh}
N_h=\{(\tilde R,\tilde v,\tilde s,\tilde u)\in\mathfrak R: \tilde R=0,\frac 1 2 (\tilde u^2+\tilde v^2)=h\}
\end{equation}
\begin{equation} \label{Dh}
D_h=\{(\tilde R,\tilde v,\tilde s,\tilde u)\in\mathfrak R: \tilde s=\pm 1, \frac 1 2(\tilde u^2+\tilde v^2)-\tilde R^a\Theta(\tilde s)^a \tilde U(\tilde s)+\tilde R^b\Theta(\tilde s)^b \tilde V(\tilde s)=h\}
\end{equation}
\[
E_h=\{(\tilde\rho,\tilde v,\tilde s,\tilde u)\in\mathfrak R :  \tilde R>0, \frac 1 2(\tilde u^2+\tilde v^2)-\tilde R^a\Theta(\tilde s)^a \tilde U(\tilde s)+\tilde R^b\Theta(\tilde s)^b \tilde V(\tilde s)=h\}.\]
The sets $N_h$  and $D_h$ are  invariant under the flow (\ref{systemh>0-2}). We call $N_h$  the {\it infinity manifold} and $D_h$ the {\it double collision manifold}. The infinity manifold  and the double collision manifold appear as a boundary submanifold glued to the energy surface $E_h$.
Note that $N_h$ is homeomorphic to a  cylinder in the three-dimensional space $(\tilde v,\tilde s,\tilde u)$.
\subsection{The flow on the infinity manifold $N_h$}

The flow on the infinity manifold is given by:
\begin{equation}
\begin{split}\label{systeminfh>0}
\tilde v'&=\Theta(\tilde s) \tilde u^2\\
\tilde s'&=\frac 1 \lambda \Theta(\tilde s)\tilde u\\
\tilde u'&=-\Theta(\tilde s)\tilde u\tilde v.
\end{split}
\end{equation}
Note first that (\ref{systeminfh>0}) the only restpoint are $(\tilde v,\tilde s,\tilde u)=(\pm \sqrt{2h},\tilde s,0)$,  $(\tilde v,\tilde s,\tilde u)=(\tilde v, 1,\pm \sqrt{2h-\tilde v^2} )$, and  
$(\tilde v,\tilde s,\tilde u)=(\tilde v, -1,\pm \sqrt{2h-\tilde v^2} )$. Note that the equilibria $(\tilde v,\tilde s,\tilde u)=(\pm \sqrt{2h},\tilde s,0)$ correspond to $1+1+1$ escape configurations. We classify the equilibrium points in the following six manifolds with boundary as follows

\[
\begin{split}
&N_h^\pm=\{(\tilde v,\tilde s,\tilde u)\in{\mathbb R}\times[-1,1]\times{\mathbb R}| (\tilde v,\tilde s,\tilde u)=(\pm \sqrt{2h},\tilde s,0)\}\\
&D_1^\pm=\{(\tilde v,\tilde s,\tilde u)\in{\mathbb R}\times[-1,1]\times{\mathbb R}| (\tilde v,\tilde s,\tilde u)=(\tilde v, 1,\pm \sqrt{2h-\tilde v^2} )\}\\
&D_{-1}^\pm=\{(\tilde v,\tilde s,\tilde u)\in{\mathbb R}\times[-1,1]\times{\mathbb R}| (\tilde v,\tilde s,\tilde u)=(\tilde v, -1,\pm \sqrt{2h-\tilde v^2} )\}
\end{split}
\]

\begin{proposition}
 The flow on $N_h$ is gradientlike with respect to the $\tilde v $ coordinate.
\end{proposition}
{\it Proof.}
We show that $\tilde v$ increases on every nonequilibrium solution.  
From the first of the equations in (\ref{systeminfh>0}) it is clear that $\tilde v'\geq 0$, since $\tilde s\in  [-1,1]$. If $ \tilde v'=0$ then either $\tilde u=0$ or $\Theta(s)=0$. In both cases we are at an equilibrium point. 
\begin{corollary}
 If we exclude the equilibrium point there are no closed or recurrent orbits on $N_h$.
\end{corollary}
{\it Proof.} If so, $\tilde v$ would vanish identically on such orbits.

As a consequence, all orbits of the flow must tend toward one of the equilibria.

\begin{figure}[t]
  \begin{center}
      {\resizebox{!}{5.5cm}{\includegraphics{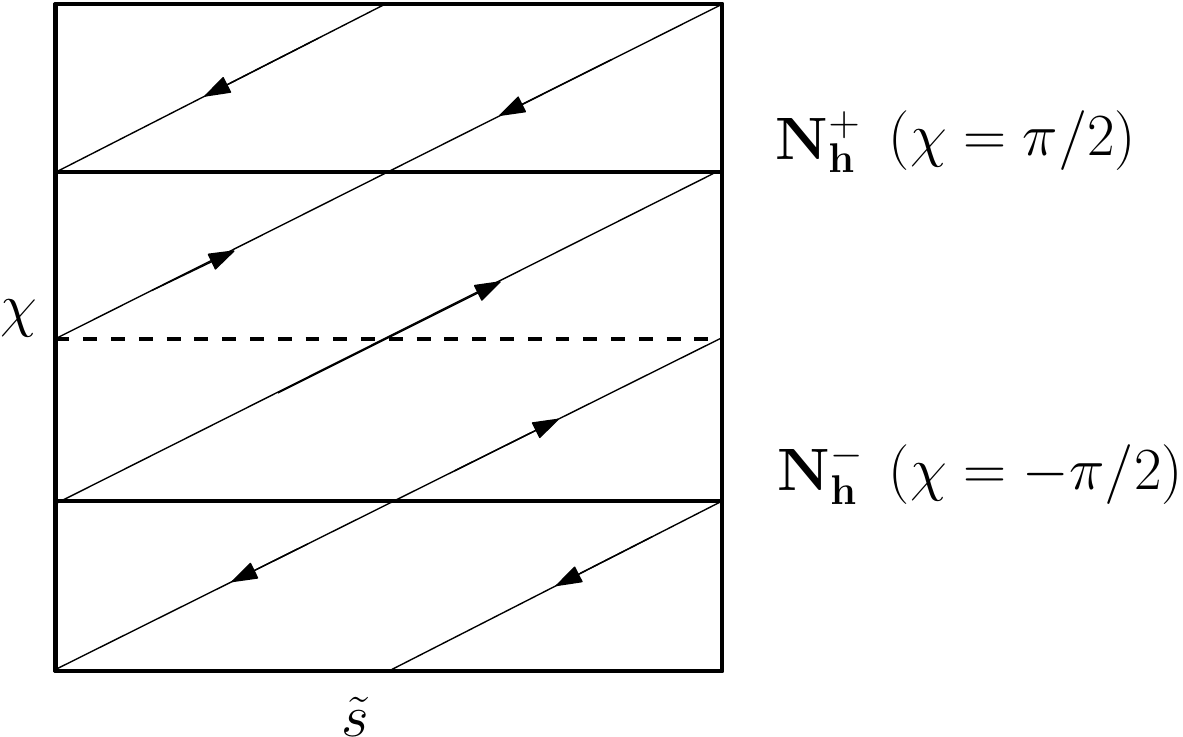}}  }
    \caption{The flow  on the  infinity manifold $N_h$ for $h>0.$\label{pos-energy}} 
  \end{center}
\end{figure}

To study the flow on $N_h$ we introduce the new  variable $\chi\in [-\pi,\pi]$ via
\begin{equation}
\begin{split}
 \tilde u &=\sqrt{2h}\cos\chi\\
\tilde v &=\sqrt{2h}\sin\chi
\end{split}
\end{equation}
we find
\begin{equation}
 \frac{d\chi}{d\tilde s}=\lambda.
\end{equation}
The behaviour of the flow changes with $\lambda$. See Figure \ref{pos-energy} for a picture of the flow for a particular value of $\lambda$


\subsection{The flow off the infinity manifold $N_h$}

We now want to study the flow off $N_h$. In order to do that we will introduce the concept of overflowing and inflowing invariant manifolds and we will then prove the existence of stable and unstable manifolds for the invariant  manifolds $N_h^\pm$ and we will find their dimension.

\begin{definition}
 Let $\phi_t$ be a smooth flow. $\bar M\equiv M\cup \partial M$ is said to be overflowing invariant  if for every $p\in \bar M$, $\phi_t(p)\in \bar M$ for all $t\leq 0$ and the corresponding vector field is pointing strictly outward and is nonzero on $\partial M$.

Let $\phi_t$ be a smooth flow. $\bar M\equiv M\cup \partial M$ is said to be inflowing invariant  if for every $p\in \bar M$, $\phi_t(p)\in \bar M$ for all $t\geq 0$ and the corresponding vector field is pointing strictly inward and is nonzero on $\partial M$.

$\bar M\equiv M\cup \partial M$ is said to be  invariant  if for every $p\in \bar M$, $\phi_t(p)\in \bar M$ for all $t$.
\end{definition}

Let $\phi_t$ be a smooth flow on a manifold C and suppose $\bar M$ is an overflowing invariant submanifold of $C$.  Let $P^u\subset TC|_{\bar M}$ be a subbundle that contains $T\bar M$ and is negatively invariant under the linearized flow $D\phi_t$. Let $I\subset P^u$ be any subbundle complementary to $T\bar M$, and let $J\subset TC$ be any subbundle complementary to $P^u$.Then   the tangent bundle to $C$ over $\bar M$ splits into three subbundles $T\bar M$, $I$ and $J$, i.e. $TC|_{\bar M}=T\bar M\oplus I\oplus J$ (where $\oplus$ is the Whitney sum).

Let $\Pi^I$, $\Pi^J$ and $\Pi^T$ be the projections onto $I,J$ and $T\bar M$ respectively corresponding to this splitting. 
We define generalized Lyapunov type numbers as follows:
For any $p\in \bar M,$ let

\beq\begin{split}
\lambda(p)&=\inf\{b\in \mathbb R^+\big|~\|\Pi^I D\phi_{-t}(p) \|b^t\rightarrow 0 \mbox{ as } \rightarrow\infty\},\\
\gamma(p)&=\inf\{a\in \mathbb R^+\big|~\|\Pi^J D\phi_t(\phi_{-t}(p)) \|/a^t\rightarrow 0 \mbox{ as } t\rightarrow\infty \},
\end{split}
\eeq
and, if $\gamma(p)<1,$ define
\begin{equation} \begin{split}
\label{defsigma} 
\sigma(p) =\inf \{s \in \mathbb{R} \,: \, D\phi_{-t}(p) \Pi^T \|& \|\Pi^J D\phi_t(\phi_{-t}(p))\|^s \rightarrow 0 \mbox{ as } t\rightarrow \infty \}.
\end{split}
\end{equation}

More computable expressions  for the Lyapunov type numbers can be derived from the expressions above, and have the following form:
\begin{align*}
\lambda(p)&=\varlimsup_{t\to \infty}\|\Pi^I D\phi_{-t}(p)\|^{1/t} \\
\gamma(p)&=\varlimsup_{t\to \infty}\|\Pi^J D\phi_t(\phi_{-t}(p))\|^{1/t} \\
\sigma(p)&=\varlimsup_{t\to \infty}\frac{\ln\|D\phi_{-t}(p)\Pi^T\|}{-\ln\|\Pi^J D\phi_t(\phi_{-t}(p))\|} 
\end{align*}
where $\|\cdot\|$ is some matrix norm. $\lambda(p),~\gamma(p)$ and $\sigma(p)$ are independent of the choice of the metric. They also have some other nice properties, see \cite{Fenichel} for more details. 

We have the following theorem:

\begin{theorem}\label{stablemanifoldtheorem}
Let $\phi_t$ be a smooth flow on a manifold C and suppose $\bar M=M\cup \partial M$ is a 
manifold with boundary overflowing invariant under $\phi_t$, with $P^u\subset TC|_{\bar M}$ a subbundle containing $\bar M$ negatively invariant under the linearized flow $D\phi_t$. Then if $\lambda(p)<1$, $\gamma(p)<1$ and $\sigma(p)\leq 0$ for all $p\in \bar M$ there exists a smooth manifold $W^u$ overflowing invariant under $\phi_t$ such that $W^u$ contains $\bar M$ and is tangent to $P^u$ along $\bar M$.

Furthermore the manifold $W^u$ is permanent under small perturbations of the flow. 
\end{theorem}
For the proof of the Theorem we refer the reader to \cite{Fenichel}.
We remark that Theorem \ref{stablemanifoldtheorem}  can be applied to inflowing invariant manifolds. In that case, the generalized Lyapunov exponents are  computed using the time reversed flow with the limits taken as $t\rightarrow -\infty $, and the phrase ''overflowing invariant`` replaced with  ''inflowing invariant``. Also $P^u$ and $W^u$ are replaced by $P^s$ and $W^s$, with $P^s$ taken to be a positively invariant subbundle under the linearized flow $D\phi_t$.

We are primarily interested in applying Theorem \ref{stablemanifoldtheorem} to the line segments  of equilibria  $N_h^\pm$. We can now prove the following:

\medskip
\begin{theorem}
The manifold $N_h^+$  has a three-dimensional local stable manifold, and  the manifold $N_h^-$ has a three-dimensional unstable manifold.
\end{theorem}
{\it Proof.~}
Let $c^\pm_{\tilde s}=(0,\pm\sqrt{2h},\tilde s,0)$ and denote by  $X:\mathfrak R\rightarrow {\mathbb R}^4$ be the vector field associated to the system (\ref{systemh>0-2}).
We want  to compute the eigenvalues of the Jacobian
\[
J: T_{x}F_h\rightarrow T_{x}F_h.
\]
where $x=(0,\tilde v,\tilde s,\tilde u)\in F_h$. In particular we want to determine the Jacobian for  $x=c^\pm_{\tilde s}$.
Note that $DX(x):{\mathbb R}^4\rightarrow{\mathbb R}^4$ leaves $T_{x}F_h$ invariant, and $J$ is the restriction of $DX(x)$ to $T_{x}F_h$.
Using system (\ref{systemh>0-2}) we have
\begin{equation}\begin{split}
 DX_1(x)(\alpha,\gamma,\sigma,\chi)&=\left(\frac{2\tilde s\tilde u}{\lambda}-(1-\tilde s^2)\tilde v\right)\alpha\\
 DX_2(x)(\alpha,\gamma,\sigma,\chi)&=(-2\tilde s\tilde u^2)\sigma+(2\tilde u \Theta(\tilde s))\gamma\\
 DX_3(x)(\alpha,\gamma,\sigma,\chi)&=(\frac 1 \lambda \Theta(\tilde s))\chi+(-\frac{2\tilde s}{\lambda} \tilde u)\sigma\\
 DX_4(x)(\alpha,\gamma,\sigma,\chi)&=(-\Theta(\tilde s)\tilde u)\gamma+(2\tilde s\tilde u\tilde v)\sigma+(-\Theta(\tilde s)\tilde v)\chi\\
\end{split}\end{equation}
where $x=(0,\tilde v,\tilde s,\tilde u)$.
From the definition (\ref{Fh}) of $F_h$ we see that
\[
T_{x}F_h=\{\nabla F_h\cdot (\alpha,\gamma,\sigma,\chi)=0\}=\{(\alpha,\gamma,\sigma,\chi)\in {\mathbb R}^4:\tilde v\gamma+\tilde u\chi=0\},
\]
and, in particular, $T_{c^\pm_{\tilde s}}F_h=\{(\alpha,\gamma,\sigma,\chi)\in {\mathbb R}^4:\pm \sqrt 2h\gamma=0\}$.

Thus, for $(\alpha,\gamma,\sigma,\chi)\in T_{x}F_h,$ we have
\[\begin{split}
J(\alpha,\gamma,\sigma,\chi)&=DX(x)(\alpha,\gamma,\sigma,\chi) \\
&=\left[\left(\frac{2\tilde s\tilde u}{\lambda}-(1-\tilde s^2)\tilde v\right)\alpha\right]{\bf e}_1+ \left[(-2\tilde s\tilde u^2)\sigma+(-2\frac{\tilde u^2}{\tilde v} \Theta(\tilde s))\chi\right]{\bf e}_2\\
&+\left[(-\frac{2\tilde s}{\lambda} \tilde u)\sigma+(\frac 1 \lambda \Theta(\tilde s))\chi\right]{\bf e}_3+
\left[(2\tilde s\tilde u\tilde v)\sigma+(\Theta(\tilde s)\frac{\tilde u^2}{\tilde v}-\Theta(\tilde s)\tilde v)\chi \right]{\bf e}_4
\end{split}
\]
where ${\bf e}_1, {\bf e}_2,{\bf e}_3,{\bf e}_4$ are the elements of the standard basis of $\mathbb R^4$.
A basis for $T_{x}F_h$ is given by the vectors ${\bf e}_1,{\bf e}_3,{\bf e}_4$
Then the matrix for $J$ in this basis is
\[\left [\begin{array}{ccc}
\left(\frac{2\tilde s\tilde u}{\lambda}-(1-\tilde s^2)\tilde v\right)~&0~&0\\
0&(-\frac{2\tilde s}{\lambda} \tilde u)~&~(\frac 1 \lambda \Theta(\tilde s))\\
0&(2\tilde s\tilde u\tilde v)\sigma~&~(\Theta(\tilde s)\frac{\tilde u^2}{\tilde v}-\Theta(\tilde s)\tilde v)\
\end{array}\right].
\]
In particular,  $J$ at $x=c^\pm_{\tilde s}$ is given by: 
\beq\left [\begin{array}{ccc}\label{Jacobian}
-\Theta(\tilde s)\tilde v~&0~&0\\
0&0~&~(\frac 1 \lambda \Theta(\tilde s))\\
0&0~&~-\Theta(\tilde s)\tilde v\
\end{array}\right].
\eeq
Thus $\bf{e}_1$ is an eigenvector with eigenvalue $\mp \Theta(\tilde s)\sqrt{2h}$.
Note that $\{\bf{e}_3,\bf{e}_4\}$ is a basis for $T_{c^\pm_{\tilde s}}N_h$. The characteristic equation for $J^\pm_{\tilde \sigma}$ restricted to $T_{c^\pm_{\tilde s}}N_h$  is
\[
-\xi(-\xi \mp\sqrt{2h}\phi(\tilde s))=0.
\]
The above equation has roots $\xi_1=0$, and $\xi_{2,3}=\mp\sqrt{2h}\Theta(\tilde s)$.
The eigenvalues of $J(c_{\tilde s})^+$ are $0$, with multiplicity one, and $-\Theta(\tilde s)\sqrt{2h}$  with multiplicity 2. The eigenvalues of $J(c_{\tilde s})^-$ are $0$, with multiplicity one, and $\Theta(\tilde s)\sqrt{2h}$  with multiplicity 2.

Let $M^\pm=\{(\tilde v,\tilde s,\tilde u)\in{\mathbb R}\times[-a,a]\times{\mathbb R}| (\tilde v,\tilde s,\tilde u)=(\pm \sqrt{2h},\tilde s,0)\}$ with $a<1$. $M^\pm$ is a proper subset of $N^\pm_h$. We now  discuss the hyperbolicity properties of $M^-$ and compute the generalized Lyapunov type numbers. Similar computations hold for $M^+$. 
Let us denote with $E^s(c_{\tilde s}^+)$ and $E^u(c_{\tilde s}^-)$ the  stable and unstable subspaces of the linearized equations. consider the following unions
\[E^s\equiv\bigcup_{\tilde s\in[-a,a]}(E^s(c_{\tilde s}^+)), \quad E^u\equiv\bigcup_{\tilde s\in[-a,a]}(E^u(c_{\tilde s}^-)). \]
Then we have
\[
TF_h|_{M^-}= E^u, \quad TF_h|_{M^+}= E^s.
\]

If we define
\[\begin{split}
P^u &= E^u\\
P^s &= E^s
\end{split}
\]
then it should be clear that $P^s$ is a positively invariant subbundle and $P^u$ is a positively invariant subbundle. 
Since   $TM^{\pm}=M^\pm\times (0,0,\mathbb{R},0)$, we can write  $TF_h|_{M-}=TM^-\oplus I\oplus J$, where $J=M^-\times (0,0,\mathbb{R},0)$ and $I\subset P^u$ (a subbundle complementary to $TM^-$).  $I$ and $J$ are the subbundles $I$ and $J$ in the geometrical set-up of  Theorem \ref{stablemanifoldtheorem}.
Hence the projections (in the basis ${\bf e_1},~\bf{e_3},~\bf{e_4}$) are 
\[
\Pi^T=\begin{bmatrix}
0&0&0\\
0&1&0\\
0&0&0
\end{bmatrix},\quad 
\Pi^I=\begin{bmatrix}
1&0&0\\
0&0&0\\
0&0&1
\end{bmatrix}, \quad 
\Pi^J=\begin{bmatrix}
0&0&0\\
0&0&0\\
0&0&0
\end{bmatrix}.
\]

Now we compute the generalized Lyapunov type numbers associated with $P^u$ in the context of Theorem \ref{stablemanifoldtheorem}. 
Using the expression of the Jacobian computed at $x=c^\pm_{\tilde s}$ (\ref{Jacobian}) we have
\[D\phi_t(c^\pm_{\tilde s})=\left [\begin{array}{ccc}
e^{-\Theta(\tilde s)\tilde vt}~&0~&0\\
0&0~&~-\frac{1}{\lambda v} e^{-\Theta(\tilde s)vt}\\
0&0~&~e^{\Theta(\tilde s)\tilde vt}
\end{array}\right].
\]
Let $\|\cdot\|$ be the matrix norm  defined by $\|A\|=(\trace(AA^t))^{1/2}$.
Then
\[
\|\Pi^I D\Phi_{-t}(c_{\tilde s}^-)\|=\trace((\Pi_ID\phi_{-t}(c^-_{\tilde s}))(\Pi_ID\phi_{-t}(c^-_{\tilde s})) ^t  )^{1/2}=\sqrt{2}e^{-\Theta(\tilde s)\sqrt{2h}t}),
\]
and hence
\[
\lambda(c_{\tilde s}^-)=\varlimsup_{t\to \infty} (\sqrt{2}e^{-\Theta(\tilde s)\sqrt{2h}t})^{1/t}=e^{-\Theta(\tilde s)\sqrt{2h}}<1
\]
for $a<1$.
A similar computation shows that, since $\phi(c_{\tilde s}^-)=c_{\tilde s}^-$, $\|\Pi^J D\phi_t(c_{\tilde s}^-)\|=0$, and thus
\[
\gamma(c_{\tilde s}^-)=\varlimsup_{t\to \infty}\|\Pi^J D\phi_t(c_{\tilde s}^-)\|^{1/t}=0<1.\\
\]
Furthermore, since $\|\Pi^J D\phi_t(c_{\tilde s}^-)\|=\|D\phi_{-t}(c_{\tilde s}^-)\Pi^T\|$=0, then  $\sigma(c_{\tilde s}^-)$ can be easily computed to be $0$ using equation (\ref{defsigma}).

At this stage we have to apply Theorem \ref{stablemanifoldtheorem}. This cannot be done directly since the vector field  is identically zero on $\partial M^-$ and thus  $M^-$ is not overflowing invariant.  This is a technical detail that can be dealt by using a $C^\infty$ bump function to modify the vector field in a small neighborhood of $\partial M^-$ as in Section 18 of \cite{Fenichel1979}, in Section  4.1 of  \cite{Wiggins} and in Proposition 2.3 of \cite{Yagasaki}. 
This concludes on the existence of a three-dimensional local unstable manifold $W_{loc}^u$.
%

Similarly one can apply a variation of Theorem \ref{stablemanifoldtheorem} valid for inflowing invariant manifolds  to prove the existence of a three-dimensional local stable manifold $W_{loc}^{s}$ for $M^+$.
This concludes the proof.

\medskip
From the theorem above we have the following:

\medskip
\begin{corollary}
 In the collinear attractive-repulsive 3-particle problem with
positive energy given by (\ref{init}), the set of initial condition leading to  $1-1-1$ asymptotic escape configuration has  positive Lebesgue measure.
\end{corollary}

 \section*{Acknowledgements}
 This work was supported by the  NSERC,  Discovery Grants Program. The first author wishes to thank Wilfrid Laurier University for their hospitality.

\begin{thebibliography}{2007}





\bibitem{Brush} S. G. Brush [1970],  
 Interatomic Forces and Gas Theory 
from Newton to Lennard-Jones, { \it Arch. Rational Mech. Anal.}, ${\bf 39},$ 1-29.


\bibitem{Fenichel}
N. Fenichel [1971],  Persistence and Smootheness of Invariant Manifolds for Flows, {\it Ind. Univ. Math. J.}, {\bf 21}, 193-225.

\bibitem{Fenichel1979}
N. Fenichel [1979], Geometric singular perturbation theory for ordinary differential equations,
{\it J. Diff. Eqns.} {\bf 31}, 53-98.

\bibitem{Frenkel} D. Frenkel and B. Smit [2002], {\it Understanding Molecular Simulation}, second edition, Academic Press , San Diago, California, USA, 2002.


\bibitem{golstein} H. Goldstein  [1980],  {\it Classical Mechanics}, Addison-Wesley, Series in Physics, Second Edition.


\bibitem{Hirsch}
M.W. Hirsch, C.C. Pugh, M. Shub [1977], {\it Invariant manifolds}, Lecture Notes in Mathematics, Vol. 583. Springer-Verlag, Berlin-New York.


\bibitem{Lewars} E. G. Lewars [2003], {\it  Computational Chemistry: Introduction to the Theory and Applications of Molecular and Quantum Mechanics}, Springer-Verlag, Berlin-New York.


\bibitem{mcgehee}
R. McGehee  [1981], Double Collisions for a classical particle system with nongravitational interactions, {\it Comment. Math. Helvetici}, $\bf 56$, 524--557.



\bibitem{Mikkola1}   S. Mikkola and J. Hietarinta [1989], A numerical investigation of the one-dimensional Newtonian three-body problem.  {\it Cel. Mech. Dyn. Astron.} ${\bf 46}$, 1Ð18. 


\bibitem{Mikkola2}   S. Mikkola and J. Hietarinta [1990], A numerical investigation of the one-dimensional Newtonian three-body problem. IIÑpositive energies.  { \it Cel. Mech. Dyn. Astron.}, ${\bf 47}$, 321Ð331. 


\bibitem{Schubart} J. Schubart [1956],  Numerische Aufscuchung periodischer Losungen im Dreikšrperproblem.  {\it Astron. Nachr.},  ${\bf 283}$, 
17Ð22. 

 \bibitem{Venturelli} A. Venturelli [2008],  A variational proof of the existence of von Schubart's orbit. {\it Discrete Contin. Dyn. Syst. Ser. B} $\bf 10$, no. 2-3, 699--717.

\bibitem{Wiggins}
S. Wiggins [1988], {\it Global Bifurcations and Chaos-Analytical Methods}, Appl. Math. Sci. 105, Springer, New York.

\bibitem{Yagasaki}
K. Yagasaki [2000], Horseshoes in Two-Degree-of-Freedom Hamiltonian Systems with Saddle-Centers, {\it Arch. Rational Mech. Anal.}, {\bf  154}, 275-296.



\end{thebibliography}
\end{document}